\documentclass[aps,prl,twocolumn,superscriptaddress,showpacs]{revtex4}
\def\be{\begin{equation}}
\def\ee{\end{equation}}
\def\bea{\begin{eqnarray}}
\def\eea{\end{eqnarray}}
\def\bi{\begin{itemize}}
\def\ei{\end{itemize}}

\usepackage{graphicx}
\usepackage{amsmath}
\usepackage{amssymb}

\begin{document}

\title{ Multi-scale Entanglement Renormalization Ansatz in Two Dimensions:\\
        Quantum Ising Model    }

\author{Lukasz Cincio}
\author{Jacek Dziarmaga}
\author{Marek M. Rams}
\affiliation{
             Institute of Physics 
             and 
             Centre for Complex Systems Research, 
             Jagiellonian University,
             Reymonta 4, 
             30-059 Krak\'ow, 
             Poland
}

\begin{abstract}

We propose a symmetric version of the multi-scale entanglement renormalization Ansatz 
(MERA) in two spatial dimensions (2D) and use this Ansatz to find 
an unknown ground state of a 2D quantum system. Results in the simple 2D quantum Ising 
model on the $8\times8$ square lattice are found to be very accurate even with the 
smallest non-trivial truncation parameter.

\pacs{ 03.67.-a, 03.65.Ud, 03.67.Hk, 02.70.-c }
\end{abstract}
\maketitle


{\bf Introduction. ---}
Over the last decade, a rapid development of efficient methods for
simulation of strongly correlated quantum systems took place, especially
in one spatial dimension (1D). It was initiated with the, by now classic,
paper of White on the density matrix renormalization group (DMRG)
algorithm \cite{White}. Recently, the subject received new acceleration
with the paper of Vidal \cite{Vidal} who proposed an elegant version of
the algorithm based on the idea that a state of a 1D quantum spin chain
can be written as a Schmidt decomposition between any two parts of the
chain. For a generic ground state, but not at a quantum critical point,
the coefficients of this Schmidt decomposition decay exponentially and the
decomposition can be truncated to a finite number of terms $d$ with
exponentially small loss of accuracy. The DMRG algoritms are equivalent to
a matrix product state Ansatz \cite{MPS} for the ground state where each
spin $S$ is assigned $2S+1$ matrices of size $d\times d$. Each matrix has
two indices to be contracted with its two nearest neighbors in 1D. The
matrix product state can be naturally generalized to two and more
dimensions by replacing the matrices with higher rank tensors to
accomodate more nearest neighbors \cite{PEPS}. These ``tensor product
states'' \cite{TPS} can also be obtained as projected entangled pair
states (PEPS) \cite{PEPS}, the latter being a more convenient
representation to prove that any quantum state can be represented
accurately by a PEPS for a sufficiently large $d$. In 2D, unlike in 1D,  
exact calculation of expectation values in a PEPS is exponentially
hard with increasing lattice size, but this problem can be overcome, at
least for open boundary conditions, by an efficient approximate method
\cite{PEPSVidal} which is linear in the system size.

The ability to make efficient and accurate zero temperature simulations in
2D is of fundamental importance for our understanding of strongly
correlated 2D quantum systems. It is enough to mention possible
applications to high-$T_c$ superconductors which effectively are 2D systems
of strongly correlated electrons on a lattice. Their, by now classic,
Hubbard model \cite{Hubbard} has not been solved exactly despite staying
in the focus of intensive research activity for several decades.

In the context of matrix product states and their generalizations, the
main difficulty is that all calculations are polynomial in the truncation
$d$, but in 2D the degree of the polynomial is too high to go far beyond
$d=2$ or $3$ which, however, may be not accurate enough. A possible
solution to this problem is the multi-scale entanglement renormalization
Ansatz (MERA) proposed in Ref. \cite{MERA} where proper
``renormalization'' of entanglement is shown to reduce the necessary $d$
by orders of magnitude. This economy of the truncation parameter $d$ was
demonstrated to be truly impressive in the 1D quantum Ising model where,
even at the critical point, a MERA with a modest $d=8$ is as accurate as a
matrix product state with $d$ in the range of a few hundreds \cite{MERA}.
Simulations with a 1D MERA are very efficient because they are polynomial
in a relatively small $d$ \cite{MERA,Montangero}.


The 1D MERA is motivated by the following real space renormalization group
algorithm. Spins on a 1D lattice can be grouped into a lattice of blocks of
two nearest neighbor spins.  There are two possible choices of block
chains, A and B, shifted with respect to each other by one lattice site.
In a decimation step of the renormalization group, each A-block is
replaced by one effective block spin whose Hilbert space is truncated to
its $d$ most important states. The most important states are the
eigenstates of an A-block's reduced density matrix with the highest
eigenvalues. However, to keep $d$ as small as possible but without loosing
much accuracy before every decimation all pairs of nearest neighbor
A-blocks are partly disentangled by $2$-spin unitary transformations
(disentanglers) acting on the $2$-spin B-blocks. The disentanglers are
optimized to minimize the entropy of entanglement of each A-block with the
rest of the lattice. They remove entanglement between those pairs of
nearest neighbor spins which belong to different A-blocks before the
A-blocks are decimated. The same basic decimation step, including
disentanglers, is then applied iteratively to the resulting decimated
lattices of block spins. It is worth mentioning, that a similar renormalization 
group was proposed earlier in Ref. \cite{CORE}, but without the disentanglers 
which are essential to keep $d$ reasonably small.

\begin{figure}[t]
\includegraphics[width=0.75\columnwidth,clip=true]{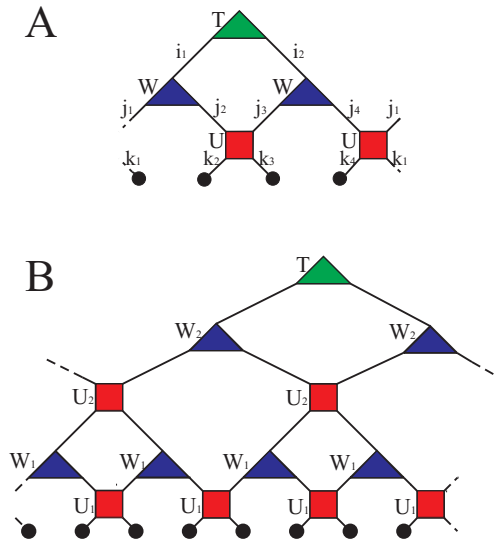}
\caption{ 
In A the 1D MERA in Eq. (\ref{1DMERA}) on a $4$-site periodic lattice of spins. 
In B the 1D MERA is generalized to a periodic lattice of $8$ spins. 
}
\label{Fig1DMERA}
\end{figure}

This renormalization group algorithm motivates the MERA. The simplest 
non-trivial example of a 1D MERA is the Ansatz for a periodic lattice 
of $N=4$ spins:
\bea
T_{i_1i_2}~~
W^{i_1}_{j_1j_2}W^{i_2}_{j_3j_4}~~
U^{j_2j_3}_{k_2k_3}U^{j_4j_1}_{k_4k_1}~~
|k_1k_2k_3k_4\rangle~.
\label{1DMERA}
\eea  
Its graphical representation is shown in panel A of Fig. \ref{Fig1DMERA}. The repeated
indices in Eq. (\ref{1DMERA}) imply summation. The lowest layer of indices $k$ number basis 
states of the $4$ spins. Here the A-blocks are the pairs of spins $(1,2)$ and $(3,4)$ and 
the B-blocks are the pairs $(2,3)$ and $(4,1)$. The $U$'s are the disentaglers; they are 
unitary matrices satisfying unitarity conditions $UU^\dagger=1$ and $U^\dagger U=1$, or 
$U^{j_1j_2}_{k_1k_2}\left(U^{j_3j_4}_{k_1k_2}\right)^*=\delta_{j_1j_3}\delta_{j_2j_4}$
and
$U^{j_1j_2}_{k_1k_2}\left(U^{j_1j_2}_{k_3k_4}\right)^*=\delta_{k_1k_3}\delta_{k_2k_4}$.
The second layer of indices $j$ numbers basis states of disentangled spins defined
by e.g. $||j_2,j_3\rangle\rangle=U^{j_2j_3}_{k_2k_3}|k_2,k_3\rangle$. The $W$'s are 
isometries or projectors which satisfy orthonormality relations 
$W^{i_1}_{j_1j_2}\left(W^{i_2}_{j_1j_2}\right)^*=\delta_{i_1i_2}$. Their job is to truncate 
the Hilbert space of the disentangled A-block spins to the $d$ most important states numbered 
by the upper indices $i\in\{1,...,d\}$: for any fixed upper index $i$, the matrix $W^i_{j_1j_2}$ 
is the $i$-th eigenstate of the A-block reduced density matrix in the basis of states of 
disentangled spins $||j_1,j_2\rangle\rangle$. The eigenstates numbered by indices $i$ become
basis states $|||i_1\rangle\rangle\rangle=W^{i_1}_{j_1j_2}||j_1,j_2\rangle\rangle$ of the
effective block spin. Finally, the top tensor $T_{i_1i_2}$, which is normalized as 
$T_{i_1i_2}\left(T_{i_1i_2}\right)^*=1$, is a quantum state in the basis $|||i_1,i_2\rangle\rangle\rangle$ 
of the effective block spins. 

In panel B of Figure \ref{Fig1DMERA}, we show a generalization of the
$4$-spin Ansatz in panel A to a periodic lattice of $N=8$ spins. The $8$
spins require one more layer of isometries and disentanglers. In general,
a lattice of $N=2^n$ spins requires $(n-1)$ layers of isometries and
disentanglers so the number of tensors that need to be stored in memory is
only logarithmic in $N$.

\begin{figure}[t]
\includegraphics[width=0.75\columnwidth,clip=true]{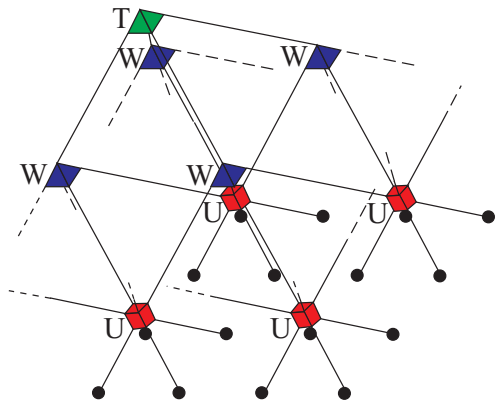}
\caption{ 
The symmetric 2D MERA on a periodic $4\times 4$ lattice. The isometries $W$ 
replace $4$-spin square plaquettes with one effective block spin in just 
one decimation step.
}
\label{ABVidal}
\end{figure}

\begin{figure}[t]
\includegraphics[width=0.55\columnwidth,clip=true]{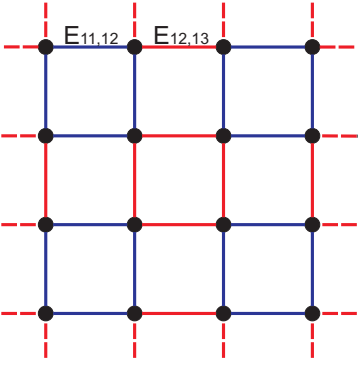}
\caption{ 
The symmetric decimation in 2D: each blue $4$-spin square plaquette is replaced
by a block spin whose Hilbert space is truncated to its $d$ most important 
states, but before this decimation a unitary $4$-spin disentangler is applied 
to each red plaquette. The disentanglers remove unwanted entanglement
between all those (red) nearest neighbor pairs of spins which belong to 
different (blue) decimation blocks.
}
\label{AB}
\end{figure}

In Ref. \cite{MERA} MERA was generalized further to 2D, and in Ref. \cite{Osborne} 
it was put in a more general unifying framework. In this paper, we propose the 
alternative 2D Ansatz in Fig. \ref{ABVidal}. In this symmetric Ansatz, $2\times2$ 
square plaquettes are replaced by effective block spins in each decimation step. 
The symmetric Ansatz is disentangling in a systematic way all those pairs of 
nearest neighbor (n.n.) spins which belong to different $2\times2$-spin decimation 
blocks, see Fig. \ref{AB} where the spins on a 2D square lattice are grouped into 
blue and red plaquettes. We propose that in each decimation step each blue 
plaquette is replaced by an effective block spin whose Hilbert space is truncated 
to its $d$ most important states, but before each decimation, the blue plaquettes 
are partly disentangled by $4$-spin unitary disentanglers acting on the red plaquettes. 
They remove entanglement between all those pairs of n.n. spins which belong to
different blue decimation blocks. Indeed, note that in Fig. \ref{AB}, all
links joining such pairs of spins are painted red. These red links are naturally 
grouped into red plaquettes and the proposed 4-spin disentanglers remove all 
the unwanted ``red'' n.n. entanglement before the following decimation. 
It is essential here that the red plaquettes are disjoint, because thanks to 
this all the unwanted ``red'' entanglement can be removed by the small $4$-spin 
disentanglers acting on individual red plaquettes. Other decimation schemes either 
do not remove all the unwanted n.n. entanglement between different decimation blocks, 
or they would require disentanglers acting on more than $4$ spins.

The symmetric variant of the renormalization group motivates MERA 
shown in Fig. \ref{ABVidal} in the case of $4\times 4$ periodic lattice. 
This graph represents the quantum state
\bea
\begin{array}{c}
T_{i_{11}i_{12}i_{21}i_{22}}
\end{array} \times \nonumber\\
\begin{array}{cc}
W^{i_{11}}_{j_{44}j_{41}j_{14}j_{11}} & 
W^{i_{12}}_{j_{42}j_{43}j_{12}j_{13}} \times \\
W^{i_{21}}_{j_{24}j_{21}j_{34}j_{31}} &
W^{i_{22}}_{j_{22}j_{23}j_{32}j_{33}} \times
\end{array}
\nonumber\\
\begin{array}{cc}
U^{j_{11}j_{12}j_{21}j_{22}}_{k_{11}k_{12}k_{21}k_{22}} &
U^{j_{13}j_{14}j_{23}j_{24}}_{k_{13}k_{14}k_{23}k_{24}} \times\\
U^{j_{31}j_{32}j_{41}j_{42}}_{k_{31}k_{32}k_{41}k_{42}} &
U^{j_{33}j_{34}j_{43}j_{44}}_{k_{33}k_{34}k_{43}k_{44}} \times
\end{array}
\nonumber\\
\left|
\begin{array}{cccc}
k_{11}&k_{12}&k_{13}&k_{14} \\
k_{21}&k_{22}&k_{23}&k_{24} \\
k_{31}&k_{32}&k_{33}&k_{34} \\
k_{41}&k_{42}&k_{43}&k_{44} 
\end{array}
\right\rangle .
\eea
Here the double subscript indices numerate rows and columns of the lattice. 
A generalization to greater $2^n\times2^n$ lattices is obtained by adding $(n-2)$ 
layers of isometries and disentanglers.


In this paper we use MERA to find the ground state of the spin-$\frac12$ transverse
quantum Ising model 
\be
H~=~
-~g~\sum_i X_i~-~ \sum_{\langle i,j\rangle} Z_i~Z_j~
\label{H}
\ee
on $2\times 2$, $4\times 4$, and $8\times 8$ periodic square lattices. Here $X$ and $Z$ 
are Pauli matrices. $T$ and all layers of different $W$ and $U$ were optimized to minimize 
total energy. Provided that the minimization preserves all constraints on $T,W,U$ 
(respectively: normalization, orthonormality and unitarity) there is no need to obtain 
$W^i$'s as leading eigenstates of reduced density matrices and to construct $U$'s as 
disentanglers that minimize the entropy of those matrices. The $W$'s and $U$'s that 
minimize the energy are at the same time good candidates for respectively the leading 
eigenstates and optimal disentanglers. 

In most calculations, we used $d=2$ in all tensors i.e. the minimal non-trivial
value of the truncation parameter, except for the $8\times 8$ lattice where it was 
necessary to increase the parameter to $d=3$, but only in the top tensor $T$
near the critical $g=3.04$. For any $g$, the initial state for the minimization was 
the Schr\"odinger cat state
$|\uparrow\uparrow\uparrow\dots\rangle+
 |\downarrow\downarrow\downarrow\dots\rangle$ which is the ground state when $g\to 0$. 
This state translates into trivial disentaglers $U=1$, the top $T$ having only two non-zero 
elements $T_{1111}=T_{2222}=1/\sqrt{2}$, and all $W$'s being non-zero only when 
$W^1_{1111}=W^2_{2222}=1$. As we were looking for the ground state, we assumed that
all tensors $T,W,U$ are real. The tensor $T$ and each tensor $W^i$ are quantum 
states on a $2\times 2$ square plaquette. Each tensor $T$ or $W^i$ has four lower indices
with each index numbering $d$ states of its corresponding spin. We assume that $T$
is symmetric under all exchanges of lower indices that correspond to symmetry 
transformations of the $2\times 2$ plaquette. As each $W^i$ is an eigenstate of a 
reduced density matrix, it must be either symmetric or anti-symmetric under each of
these symmetry transformations. In all considered cases, we found that the lowest energy 
is obtained when all $W^i$'s are assumed symmetric under all transformations. In this 
symmetric subspace it is convenient to parametrize the tensors as (here $d=2$)
\bea
T_{abcd} &\simeq& 
\sum_{\alpha=1}^6 
t_\alpha ~ v^\alpha_{abcd}~, \nonumber\\
W^i_{abcd} &\simeq& 
\sum_{\alpha=1}^6 
w^i_\alpha ~ v^\alpha_{abcd}~, \nonumber\\
U &=& 
\exp\left( i \sum_{\alpha=1}^{21} q^\alpha A_\alpha\right )~,
\label{q's}
\eea
where $\simeq$ means equality up to normalization. Here 
$v^\alpha_{abcd}=\langle abcd|v^\alpha\rangle$ where the states
\bea
|v^1\rangle&=&|0000\rangle,~~|v^2\rangle=|1111\rangle, \nonumber\\
|v^3\rangle&=&
\left(|0110\rangle+|1001\rangle\right)/\sqrt{2}, \nonumber\\
|v^4\rangle&=&
\left(|1000\rangle+|0100\rangle+|0010\rangle+|0001\rangle\right)/2, \nonumber\\
|v^5\rangle&=&
\left(|0111\rangle+|1011\rangle+|1101\rangle+|1110\rangle\right)/2, \nonumber\\
|v^6\rangle&=&
\left(|1100\rangle+|0011\rangle+|0101\rangle+|1010\rangle\right)/2,  
\eea
are a basis of symmetric states on the $2\times2$ plaquette. $A_\alpha$'s are imaginary 
$4$-spin hermitian operators invariant under the symmetries of the $2\times 2$ plaquette:
\bea
A_1 &\simeq& Y_1+Y_2+Y_3+Y_4              ~,   \nonumber\\
A_2 &\simeq& X_1Y_2+Y_1X_2+X_2Y_4+Y_2X_4+      \nonumber\\
    &      & X_3Y_4+Y_3X_4+X_3Y_1+Y_3X_1~,..., \nonumber\\
A_{21} &\simeq& Y_1Y_2Y_4+Y_2Y_4Y_3+Y_4Y_3Y_1+Y_3Y_1Y_2~.
\eea 
Each $A_\alpha$ is a symmetrized sum of tensor products of Pauli matrices with each 
term in the sum including an odd number of $Y$'s. They are normalized so that 
${\rm Tr}A^\dag_\alpha A_\beta=\delta_{\alpha\beta}$.

\begin{figure}[t]
\includegraphics[width=0.99\columnwidth,clip=true]{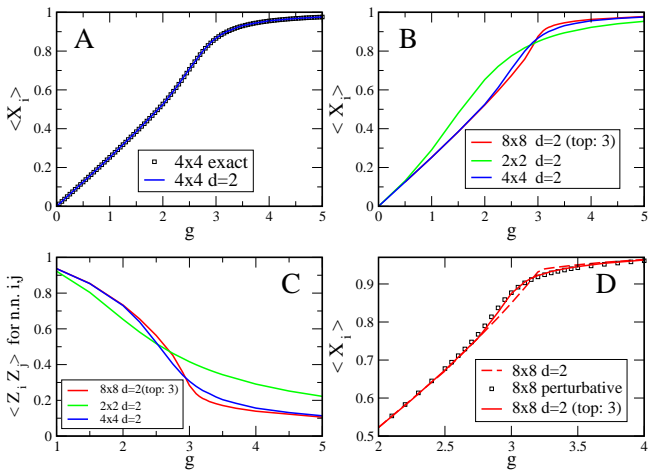}
\caption{ 
In A, we compare transversal magnetization $\langle X \rangle$ on the $4\times4$ 
lattice obtained from MERA and exact diagonalization. In B and C, transversal 
magnetization and nearest neighbor ferromagnetic correlator obtained from MERA 
are shown for different lattice sizes. In D, we compare transversal magnetization 
on the $8\times8$ lattice when $d=2$ in all tensors and when it is increased to 
$d=3$ in the top tensor with the perturbative results from Ref. \cite{Sandvik}.
}
\label{Fig_MERA_results}
\end{figure}

The minimized energy is a sum of all bond energies 
$
E_{i,j}=
\langle
-\frac14gX_i-\frac14gX_j-Z_iZ_j
\rangle~.
$
However, thanks to the assumed symmetry of the tensors, only some of them need to be 
calculated. For example, on the $4\times 4$ lattice in Fig. \ref{AB}, one needs to 
evaluate only two bond energies: $E_{11,12}$ and $E_{12,13}$. By symmetry, all
other bond energies are equal either $E_{11,12}$ or $E_{12,13}$ and the total
energy is $\langle H \rangle=16E_{11,12}+16E_{12,13}$. In a similar way, the
$8\times 8$ square lattice has $6$ and, in general, an $N\times N$ square lattice 
has $\frac{N^2}{16}+\frac{N}{4}$ independent bond energies. The total number of bonds 
is $2N^2$ so for a large $N$ we save a factor of $32$ simply by using the assumed tensor 
symmetries. Thus for a large $N$, the cost of calculating energy is proportional to the 
lattice size times the cost of calculating any bond energy $E_{i,j}$ which is logarithmic 
in $N$ and polynomial in $d$. Here the proof follows similar lines as in Ref. \cite{MERA}.
Indices are contracted along causal cones whose horizontal cross-section is $3\times3$ 
(or $4\times4$) spins when cut above (or below) a layer of isometries $W$. To avoid the 
intermediate $4\times4$ stage we do not apply all isometries first and then all 
disentanglers, but we apply some isometries earlier than other gradually including 
disentanglers i.e. we pass through a series of intermediate non-horizontal 
cross-sections never exceeding 11 spins. 
    
Energy was minimized with respect to the variational parameters 
$\{t_\alpha,w^i_\alpha,q^\alpha\}$ in Eq. (\ref{q's}) using different standard 
minimization routines, but the best performance was achieved with the simplest steepest 
descent method with gradients of the energy estimated from finite differences. 
Our calculations demonstrate that energy of MERA can be minimized in a fairly 
straightforward manner. 
  
In figure \ref{Fig_MERA_results}, we summarize our results for $2\times 2$, $4\times 4$,
and $8\times 8$ periodic square lattices. Of special interest are panels A and D, where 
we compare transversal magnetization obtained from MERA with exact results on the 
$4\times4$ lattice and perturbative results on the $8\times8$ lattice. On the $4\times4$
lattice $d=2$ was accurate enough, but on the $8\times8$ lattice $d$ in the top
tensor had to be increased to $d=3$. This was necessary because with increasing
lattice size Ising model develops a critical point at $g=3.04$ - this tendency
can be seen in panels B and C.


In conclusion, we proposed and tested a symmetric version of MERA in 2D. Using the 
smallest non-trivial truncation parameter $d=2$ in most tensors and fairly 
straightforward optimization methods
we obtained surprisingly accurate numerical results for the ground state of the 2D 
quantum Ising model. This is, we think, an encouraging result but, as the Ising model 
that we consider is relatively simple, it remains to be seen how well MERA can deal with 
more complicated models.

{\it Note added. } When this paper was in the final stage of preparation, the e-print 
\cite{symVidal} appeared where similar symmetric Ansatz was proposed.


{\it Acknowledgements.} We are indebted to Bogdan Damski and Anders Sandvik for 
providing us with exact results. Discussions with Bogdan Damski, Maciek Lewenstein, 
and Kuba Zakrzewski are appreciated. This work was supported in part by Polish 
government scientific funds (2005-2008) as a research project and in part by Marie 
Curie ATK project COCOS (contract MTKD-CT-2004-517186).



\begin{thebibliography}{99}

\bibitem{White} S. R. White, 
                Phys. Rev. Lett. {\bf 69}, 2863 (1992). 

\bibitem{Vidal} G. Vidal, 
                Phys. Rev. Lett. {\bf 91}, 147902 (2003);  
                Phys. Rev. Lett. {\bf 93}, 040502 (2004); 
                Phys. Rev. Lett. {\bf 98}, 070201 (2007).

\bibitem{MPS} S. Rommer and S. \"Ostlund, 
              Phys. Rev. B {\bf 55}, 2164 (1997);
              M.-C. Chung and I. Peschel,
              Phys. Rev. B 62, 4191 (2000).

\bibitem{PEPS} F. Verstraete and J. I. Cirac, 
               arXiv:cond-mat/0407066 (2004);
               F. Verstraete, M. M. Wolf, D. Perez-Garcia, and J. I. Cirac,
               Phys. Rev. Lett. {\bf 96}, 220601 (2006);
               V. Murg, F. Verstraete, and J. I. Cirac,
               Phys. Rev. A {\bf 75}, 033605 (2007);
               N. Schuch, M. M. Wolf, F. Verstraete, and J. I. Cirac,
               Phys. Rev. Lett. {\bf 98}, 140506 (2007);
               N. Schuch, M. M. Wolf, F. Verstraete, and J. I. Cirac,
               arXiv:0708.1567.

\bibitem{TPS} T. Nishino, K. Okunishi, Y. Hieida, N. Maeshima, and Y. Akutsu,
              Nucl. Phys. B {\bf 575}, 504 (2000).

\bibitem{PEPSVidal} J. Jordan, R. Orus, G. Vidal, F. Verstraete, J. I. Cirac,
                    arXiv:cond-mat/0703788;
                    A. Isacsson and O. F. Syljuasen,
                    arXiv:cond-mat/0604134.

\bibitem{Hubbard} J. Hubbard, 
                  Proc.Roy.Soc.(London) A {\bf 272}, 238 (1963);
                  {\it ibid.} {\bf 281}, 401 (1964). 

\bibitem{MERA} G. Vidal, 
               arXiv:cond-mat/0512165 (to appear in Phys. Rev. Lett.); 
               arXiv:0707.1454.

\bibitem{CORE} C. J. Morningstar and M. Weinstein, 
               Phys. Rev. D {\bf 54}, 4131 (1996).  

\bibitem{Montangero} M. Rizzi and S. Montangero, 
                     arXiv:0706.0868.

\bibitem{Osborne} C. M. Dawson, J. Eisert, and T. J. Osborne, 
                  arXiv:0705.3456.

\bibitem{Sandvik} A. W. Sandvik, 
                  Phys. Rev. E {\bf 68}, 056701 (2003).

\bibitem{symVidal} G. Evenbly and G. Vidal, 
                   arXiv:0710.0692.

\end{thebibliography}
\end{document}